\documentstyle[prl,aps,epsf]{revtex}
\input epsf

\begin{document}

\twocolumn[\hsize\textwidth\columnwidth\hsize\csname @twocolumnfalse\endcsname
\draft
\title{Ratchet Cellular Automata}

\author{ M. B. Hastings$^1$, C. J. Olson Reichhardt$^{2}$, 
and C. Reichhardt$^1$}
\address{$^1$Center for Nonlinear Studies and $^2$T-12, 
Theoretical Division, Los Alamos
National Laboratory,  Los Alamos, New Mexico 87545 
}
\date{\today}
\maketitle
\begin{abstract}
In this work we propose a ratchet effect which provides a general
means of performing clocked logic operations on discrete particles,
such as single electrons or vortices.  The states are
propagated through the device by the use of an applied AC drive.
We numerically demonstrate that
a complete logic architecture is realizable using this ratchet.  We
consider specific nanostructured superconducting geometries using
superconducting materials under an applied magnetic field, with the 
positions of the individual vortices in samples acting as the logic states.  
These devices can be used as the building blocks 
for an alternative microelectronic architecture.
\end{abstract}
\pacs{PACS: 74.60.Ge,05.70.Ln,05.40.-a}
]
As the size scale for microelectronics continues to decrease,
limits to the efficiency of standard architectures 
will at some point be exhausted 
which will mandate the necessity  of switching to 
alternative device architectures\cite{CZ97}.  Any such architecture requires a
means of storing state information, as well as a means of performing
logic operations in a clocked fashion on the state information.

In this work, we propose a means of performing logic operations based on
a novel deterministic ratchet mechanism.
This provides a general means of computing
with discrete particles,
be they 
vortices in superconductors, 
single electrons in coupled quantum dots, Josephson
vortices, or ions, via a
non-equilibrium drive applied to the system.  

The use of single electron charges for storing state information has
been well studied.
An example of this is the quantum dot cellular
automata (QCA)\cite{Lent,Amlani} where the positions of the electrons are used
to create the logic states and 
adiabatic changes are performed
to the system Hamiltonian, so that logic operations are performed with
the system always remaining in its ground state.
A magnetic version of the QCA has also been proposed \cite{Cowburn}.
One disadvantage of QCA 
is that it is currently   
limited to operation at very low temperatures.
Further, the need to perform adiabatic changes to the Hamiltonian 
limits the processing speed.  The ratchet mechanism discussed in this
paper provides a means of significantly increasing this speed.

A different approach to storing information
is to use superconducting nanostructured 
arrays in a magnetic field where positions of
the vortices define the logic state\cite{Puig}.  This is the specific system
we consider in this paper 
to
numerically demonstrate the feasibility of
our ratchet.
When a magnetic field is applied to a superconductor, the
flux enters in the form of individual quantized vortices
which repel each other and form a triangular lattice.  
Recent work on mesoscale superconductors has demonstrated that 
individual vortices can be captured in a single sample 
\cite{Geim,Peeters,Reichhardt}.
Additionally, several groups have nanostructured the surface of a 
superconductor with pinning sites which act as areas that
capture vortices \cite{Baert,Harada,Schuller}. 
These nanostructured arrays are made by 
creating magnetic or non-magnetic dots, and the dot geometry 
of the individual dots can be controlled. 
Consider two parallel elongated dots or pinning sites with the
elongation in plane and a magnetic field perpendicular to the plane
at a strength such that each dot captures exactly one vortex. 
If the dots are in close proximity, then the positions of the 
vortices in the two dots will be correlated due to their mutual repulsion. 
The vortices will be arranged such that one vortex is located 
at the top of the dot and the other at the bottom of the adjacent 
dot. The state with the vortex at the top we consider to be
a logic value of $1$, while
the state where the vortex is at the bottom of the dot is a logic $0$.
Experiments and simulations
on small $2\times2$ superconducting arrays have observed
such states\cite{Puig,ccj}.    

\begin{figure}
\epsfxsize=2.2in
\epsfbox{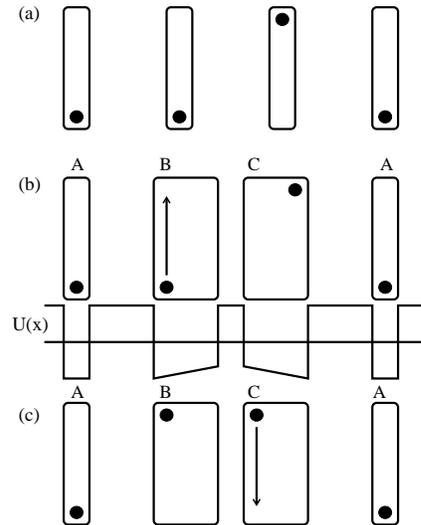}
\caption{(a)  Pipeline without ratchet.  (b)  Schematic of
ratchet for pipeline.  $U(x)$ indicates potential as a function of $x$.
Positions and motion of vortices are indicated at ${\bf J}=0$.  (c)
Schematic of second stage of ratchet, for ${\bf J}=-J{\hat y}$.}
\end{figure}

In order to create logic devices based on these dots, we require
a mechanism for propagating a flip or change of logic state through the
dot array, as well as to perform logic operations.  
Further, the new state must propagate
at a constant rate so that a constant clock speed 
can be achieved.  For the simplest geometry of Fig.~1(a), alternate vortices
prefer to align in opposite logic states due to repulsion
between the vortices.  We have indicated a defect
at the left of this configuration, and ideally would like this defect
to propagate to the right by flipping vortices, 
carrying new state information.  However,
since the vortex system is dissipative, moving the defect requires thermal
activation and is hence slow and equally likely to occur in either
direction.
To remedy this, we propose a means of propagating information entirely
distinct from that in the QCA system: in a
suitable dot geometry, 
deterministic mechanical ratchet effects can be used to drive the
system from one state to the next\cite{Bar}.  The dissipative nature of the
dynamics increases the speed of the device,
while the short distance traveled by the vortices
in the device limits the power consumption to very low levels.
To further increase the speed of the device, the same
ratchet effect can be employed with different building blocks, including
Josephson vortices and electrons in quantum dots, as we discuss below.

To describe the ratchet, we recall that
a vortex is well modeled as an overdamped
particle so that the velocity
 ${\bf v}$ 
is proportional to the
net force on it, where the force arises from vortex-vortex
interaction, confining potential, and Lorentz forces due to
any applied external current.  
The equation of motion is then 
\begin{equation}
{\bf F} = {\eta} {\bf v}  = {\bf F}_{vv} + {\bf F_{s}} + {\bf F}_{ac} 
\end{equation}
where $\eta$ is the viscous damping coefficient and 
${\bf F}_{vv}$ is the repulsive vortex-vortex interaction.
The force from the confining potential
is ${\bf F}_{s} = - \nabla U(x,y){\hat {\bf r}}$, and 
${\bf F}_{ac}$ is the Lorentz force from an applied ac current ${\bf J}$,
${\bf F}_{ac} = {\bf J(t)}\times {\hat {\bf h}}\Phi_{0}d$, where
$\Phi_{0} = 2.07\times 10^{-15}$ T m$^{2}$ is the elementary flux quantum of
one vortex and $d$ is the sample thickness.
The applied $H$ is out of the
plane in Fig.~1.
To drive the ratchet, we apply a current, uniform across the sample, 
in the $\hat y$-direction, producing
a Lorentz force that moves the vortices in the $\hat x$-direction.  To
obtain a uniform current, it may be necessary to attach multiple leads
to the sample to deal with sample inhomogeneities.
Additional current in the $\hat x$ direction can be
applied locally to change the state of individual vortices and
to enable input to the vortex logic.

In Fig.~1(b) we show the basic ratchet mechanism to propagate logic
information along a pipeline.  There are three different types of
potential wells, labeled by letters A, B, and C and alternating
in that pattern.  The A
wells are narrow, so that vortices in these wells have
little ability to move in the $\hat x$-direction.  The B and C
wells are both wider, with the B well having an overall tilt in
the potential to the left side of the well and the C well having a tilt to
the right side.  
The form of $U(x)$ is illustrated below Fig.~1(b), where we plot
$U(x)$ along a line passing through the center of the wells.
To drive the ratchet, 
the external field $J(t)$ is driven through a series of states,
${\bf J}=0, {\bf J}=-J {\hat y}$, and ${\bf J}=+J{\hat y}$, consecutively. 
The external current is taken
sufficiently strong so that for ${\bf J}=\mp J{\hat y}$ 
it can push the vortex to the left or right side, respectively, of
the well, overcoming the tilt in $U(x)$ in the B and C wells, while
at ${\bf J}=0$, the tilt in the B and C wells determines the $x$-position
of the vortices.  Thus, the spacing between the vortices changes as
${\bf J}$ changes, with a narrow spacing between some and a
wide spacing between others,
altering the strength of the interaction between different
neighbors.  The lowest energy state for the system is to put the
defect between vortices which have the furthest spacing.
For ${\bf J}=0$, this is between wells B and C; for ${\bf J}=-J{\hat y}$ 
it is between wells C and A; and for ${\bf J}=+J {\hat y}$ it is 
between wells A
and B.  Thus, if the left-most vortex is held fixed in Fig.~1(b),
at $J=0$, the vortex in the B well moves upwards as indicated by
the arrow.  Then ${\bf J}$ is switched to $-J{\hat y}$ and the vortices move
as indicated in Fig.~1(c).  Finally, switching ${\bf J}$ to $+J{\hat y}$
moves the defect one more position to the right, and the ratchet can repeat.
In this process, the alternating current raises
the energy of the system by moving the vortices; this energy is then
dissipated as the vortex moves.  However, there is still
an energy barrier to the vortex motion.  As the vortex in the
B well moves upward in Fig.~1(b), its energy initially increases,
before dropping as the vortex completes its motion. 
By adding a suitable additional potential $U(y)\propto y^2$, we are able
to remove this barrier, changing the thermal ratchet into a deterministic
ratchet.

We demonstrate numerically the operation of these devices
via simulation.  The optimal geometry of the dots has
the ratio between the wide and narrow horizontal spacings between vortices
equal to $2$.  To include a realistic 
finite separation between dots, we
considered ratios of approximately $1.3-1.5$.  Larger ratios
increase the speed and ease of design of the device.
We consider two types of vortex-vortex
interactions.  The first, appropriate for bulk samples, is 
${\bf F}_{vv} = (\Phi_0^2 d/2\pi\mu_0\lambda^3) K_{1}(r/\lambda){\hat {\bf r}}$, 
where $K_{1}(r/\lambda)$ 
is the modified Bessel function that falls off monotonically 
with $r$, and $\lambda$ is the London penetration depth.  
The second form we consider,
appropriate for a thin film superconductor, is 
${\bf F}_{vv} = (\Phi_0^2/\mu_0 \pi \Lambda){\hat {\bf r}}/r$, 
where $\Lambda$ is the thin film screening length\cite{clem2}.

In Fig.~2(b) we show the results from a simulation of a pipe line, 
with a geometry where the flip can be seen to propagate linearly in
time with the AC drive. Fig.~2(a) shows the thermal ratchet for the case
without the additional potential $U(y)\propto y^2$, indicating occasional
reverse steps.
In these simulations we considered a pattern of 144 wells with a
repeat pattern length of $5\lambda$, thin well diameter
$0.48\lambda$, and wide well diameter  $1\lambda$.
The close spacing was $1.5\lambda$ and the far spacing was $2\lambda$.
The length 

\begin{figure}[!t]
\epsfxsize=3.4in
\epsfbox{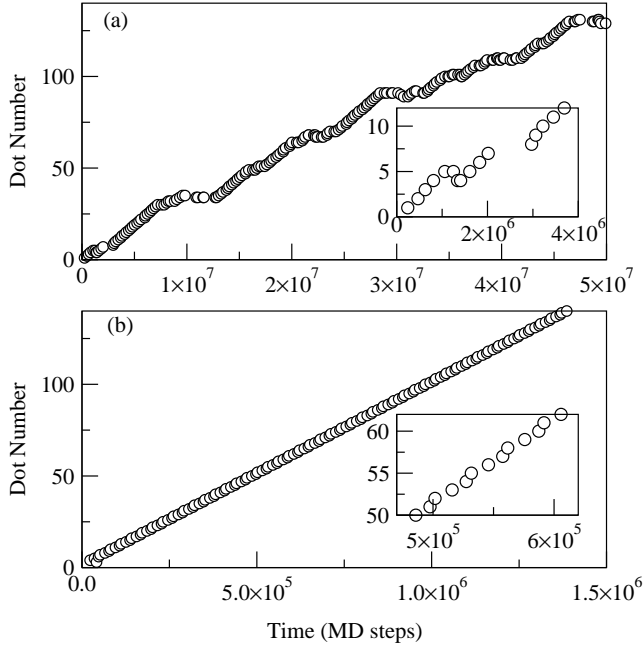}
\caption{Simulated signal propagation through a pipeline.  The time at which
the vortex in each well changes states is indicated.  (a) Thermally
activated ratchet operating at $T=0.5$.  
Inset: Detail of the occasional backwards motion
of the signal, also showing time periods when the signal does not
propagate forward.  (b) Deterministic ratchet operating at $T=0.$
The signal is perfectly clocked.  Inset: Detail showing the slight
asymmetry in switching times of the three well shapes.}
\end{figure}

\hspace{-13pt}
of the wells in the transverse direction, not counting the
confining ends, was $1.2\lambda$.  The simulation illustrated in 
Fig.~2(b) required 10000 molecular dynamics
steps to move the signal over by three wells.

In terms of real material parameters\cite{param,mgb2}, 
the operating frequency can be
written as $\nu = 3/(dt \tau)$, with the simulation time unit
$\tau = \mu_0 \lambda^{3} d/(\xi^{2} \rho_N)$, where
$\mu_0$ is the permeability of free space,
$d$ is the film thickness, which we assume to be $d=200$ nm, and
the London penetration depth $\lambda$ for selected materials is:
YBa$_2$Cu$_3$O$_{7-\delta}$ (YBCO), $\lambda = 156$ nm;
Bi$_2$Sr$_2$CaCu$_2$O$_8$ (BSCCO), $\lambda = 250$ nm;
MgB$_2$, $\lambda = 85$ to 203 nm.
The resulting frequencies are:
YBa$_2$Cu$_3$O$_{7-\delta}$, $\nu = 160.2$ MHz;
Bi$_2$Sr$_2$CaCu$_2$O$_8$, $\nu = 86.7$ MHz;
MgB$_2$, using midpoint values: $\nu = 315$ MHz.

The frequencies above are for non-optimized well geometries, chosen
instead to be readily manufactured using present day technology.  Particularly
in BSCCO, the average spacing between wells is 0.5$\mu$m; much smaller
structures than this could be created, which would have higher operating
speeds due to the larger vortex-vortex interaction forces.  The maximum
operating frequency of the vortex cellular automaton is set by 
the depairing frequency of the Cooper pair in the BCS materials.
In Nb, which has a gap of $\Delta = 1.55$ meV, the depairing 

\begin{figure}[!t]
\epsfxsize=3in
\epsfbox{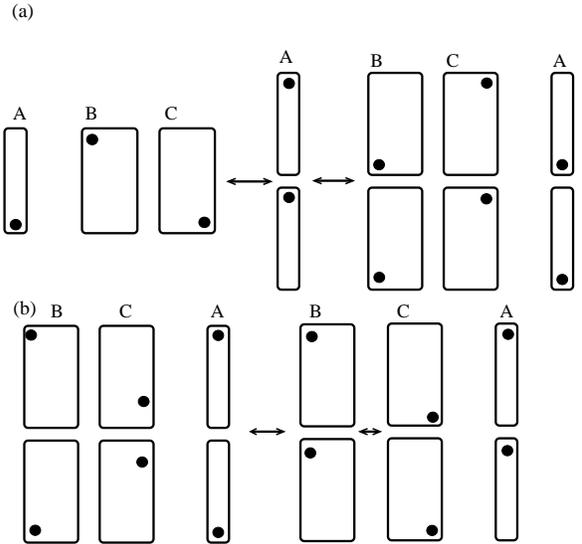}
\caption{(a)  Schematic of fanout.  The spacing between wells is slightly
increased where marked by the double arrow.  (b)  Schematic of NAND gate.
Spacing between wells is slightly increased where marked by the double
arrow, to force both pipelines into the same state.}
\end{figure}

\hspace{-13pt}
frequency 
is 688 GHz.  YBCO, with a gap of $\Delta=20$ meV, and BSCCO, with a
gap of $\Delta = 38$ meV, have depairing frequencies in the terahertz
range.  Measured gaps for MgB$_2$ range from $\Delta$=1.8 to 
7.5 meV.  We can ignore the skin effect, as for the penetration depth 
in these materials the skin effect is irrelevant for frequencies below 
$\sim 10$ THz.

The dissipation of the device due to motion of the vortices
is negligible; the resulting energy to switch a single cell is of order 
$10^{-17}$ J for MgB$_2$.  There is an additional 
surface power dissipation \cite{Clem}, which is also small, of order
$10^{-15}$ J per cycle or less.  These small dissipation energies
are similar to those for magnetic QCA devices \cite{Cowburn}.

To realize a pipeline which propagates signals in reverse, the
A well in Fig.~1(b) can be replaced with a wider well centered on the
same point, with a potential such that the vortex is in the center of the
well for ${\bf J}=0$, and moved to the sides of the well for ${\bf J}\neq 0$.
In order to make a complete logic architecture 
the basic units also include a fanout and
a NAND gate, which are illustrated in Figs.~3(a-b).
The exit from the fanout consists of two pipelines.  Due to repulsion
between particles in the neighboring pipeline, an additional potential 
which biases the vortices towards the bottom of the wells is
added to the wells in the top pipeline, and conversely for the wells in
the bottom pipeline.  The well spacing at the fanout
itself is slightly increased to enable the C cell immediately to the
left of the fan to
respond to its left neighbor rather than its two right neighbors.
The spacing between the next A and B wells is also slightly increased.  
In the NAND gate, the
pipelines before the gate have a narrower horizontal spacing.  This
increases the coupling between successive cells within a given pipeline,
compared to that between pipelines, enabling distinct
signals to propagate in each pipeline.  At and after the
gate, the horizontal spacing is increased, so that both pipelines
must be in the same state.  A slight upward bias is applied to the vortices
in the gating cells (the center A and B cells) 
as shown, to give the system a preferred state
if the inputs are in an opposite state, realizing a NAND gate.  By taking
a large number of neighboring pipelines and varying the spacing in selected
places, a very compact design for a gate array can be constructed.
The basic ratchet automatically includes an inverter, since neighboring
vortices assume opposite logic states.  An XOR and wire crossing can be
realized using the above devices as basic components.
We have also performed simulations confirming the device
geometries illustrated in Fig.~3\cite{web}.

While we propose this ratchet effect in the context
of information storage using vortex position, the ratchet effect can be
generalized to other systems, including the case of electron
charges in a quantum dot.  In this case, the maximum operating frequency
is set by the level spacing of the dot.  To obtain faster operating
speed using vortices, the vortices can be replaced with Josephson vortices.
In the ratchet described above, the vortices move along the boundaries of
the cells, either vertically along the side boundary or horizontally along
the top.  By placing a thin strip of insulating material around the border of
the cells in Fig.~1, the vortex core will exist only in the insulating material.
Since there is no normal core, the dissipation $\eta$ is greatly reduced and
the speed increased.  We anticipate that this will enable the very low
dissipation discussed above to be combined with high speed.
Our system may also be physically realizable for 
ions in dissipative optical light arrays 
where the ion motion is damped and the potentials
can be tailored by adjusting the optical landscape \cite{Optical}. 
A variation of this system could also be constructed using charged 
colloidal particles in optical trap arrays
\cite{Bechinger},
where the colloids can be driven with an AC fluid flow, electric field,
or by oscillating the trap.

In summary, we have shown that in order
to perform clocked computations on a classical system, it is necessary
to drive the system out of equilibrium.  Ratchets are a fundamental
aspect of non-equilibrium statistical physics that have been much studied
in recent years.  We have proposed a practical application of 
a ratchet mechanism to produce clocked logic operations for 
discrete particles by using an applied AC drive.  With numerical
simulations we have shown that a complete logic architecture can be
realized. We have specifically demonstrated this mechanism for
vortices in superconducting geometries. Our results 
should be generalizable for other systems  
such as single electrons in quantum dots, Josephson vortices, and
ions in optical traps.  

{\it Acknowledgments---}
We thank B. Janko for initial inspiration for this work, and 
W. Kwok and T. A. Witten for useful discussions.
This work was supported by the US DOE under Contract No. W-7405-ENG-36.

	\end{document}